\documentclass[aps,prl,twocolumn,superscriptaddress,showpacs]{revtex4}

\usepackage{graphicx} 
\begin{document}

\title{Quantum spin correlations in an organometallic alternating sign chain}

\author{M. B. Stone}
 \affiliation{Neutron Scattering Science Division, Oak Ridge National Laboratory, Oak Ridge, Tennessee 37831 USA}

\author{W. Tian}
 \affiliation{Department of Physics and Astronomy, The University of Tennessee,
Knoxville, Tennessee 37996, USA}
 \affiliation{Ames Laboratory and Department of Physics and Astronomy,
 Iowa State University, Ames, Iowa 50011 USA}

\author{M. D. Lumsden}
 \affiliation{Neutron Scattering Science Division, Oak Ridge National Laboratory, Oak Ridge, Tennessee 37831 USA}
\author{G. E. Granroth}
 \affiliation{Neutron Scattering Science Division, Oak Ridge National Laboratory, Oak Ridge, Tennessee 37831 USA}
\author{D. Mandrus}
 \affiliation{Materials Science and Technology Division, Oak Ridge National Laboratory,
Oak Ridge, Tennessee 37831, USA}

\author{J.-H. Chung}
\affiliation{NCNR, National Institute of Standards and Technology, Gaithersburg, Maryland 20899, USA}
\affiliation{Department of Materials Science and Engineering, University of Maryland, College Park, Maryland 20742, USA
}

\author{N. Harrison}
\affiliation{National High Magnetic Field Laboratory, LANL, Los Alamos, New Mexico 87545}

\author{S. E. Nagler}
 \affiliation{Neutron Scattering Science Division, Oak Ridge National Laboratory, Oak Ridge, Tennessee 37831 USA}

\begin{abstract}
High resolution inelastic neutron scattering is used to study excitations in the organometallic magnet DMACuCl$_3$.  The correct magnetic Hamiltonian describing this material has been debated for many years.  Combined with high field bulk magnetization and susceptibility studies, the new results imply that DMACuCl$_3$ is a realization of the $S=1/2$ alternating antiferromagnetic-ferromagnetic (AFM-FM) chain.
Coupled-cluster calculations are used to derive exchange parameters, showing that the AFM and FM interactions have nearly the same strength.  Analysis of the scattering intensities shows clear evidence for inter-dimer spin correlations, in contrast to existing results for conventional alternating chains.  The results are discussed in the context of recent ideas concerning quantum entanglement.

\end{abstract}


\maketitle

Low dimensional, dimerized quantum spin systems exhibit diverse physics ranging from simple spin gaps to exotic spin liquids\cite{turnbullsummary,stonephccprb,zheludevIpac,poilblanc2006}.  A prototypical example is the $S=1/2$ Heisenberg alternating chain (HAC), defined by the Hamiltonian ${\mathcal H} = 
\sum_{n} J_{1}{\mathbf S}_{2n-1}\cdot {\mathbf S}_{2n}
+J_{2} {\mathbf S}_{2n}\cdot {\mathbf S}_{2n+1}$.  The HAC has been widely studied theoretically \cite{abharris,uhrigschulz,barnesafmchain} and experimentally
\cite{gxucoppernitrateprl,blake1997}.  Considering $J_{1} > 0$ antiferromagnetic (AFM), ground state behavior of the HAC depends on the ratio $\alpha = J_{2}/J_{1}$.  $\alpha = 1$ corresponds to the quantum critical pure AFM chain.   When $|{\alpha}| << 1$ the system behaves as nearly independent dimers, and perturbation
theory in $\alpha$ provides a straightforward approach to calculate physical properties of the spin gap system.  For $\alpha < 0$, one has a realization of the antiferromagnetic-ferromagnetic (AFM-FM) HAC.  For $\alpha << -1$, the HAC maps onto the $S=1$ AFM Haldane chain with the dimer viewed as a composite object \cite{HidaPRBs1992,watanbe1999,zheng2006}.  The regime $\alpha \approx -1$ is particularly interesting since the system exhibits intermediate behavior between dimer and Haldane physics \cite{hidabocquet,HidaPRBs1992}.  Calculating physical properties must be done carefully since perturbation theory from the dimer limit is not a good approximation here.

%

\begin{figure}
\centering\includegraphics[scale=0.85,clip=true,angle=-90,keepaspectratio=true ]{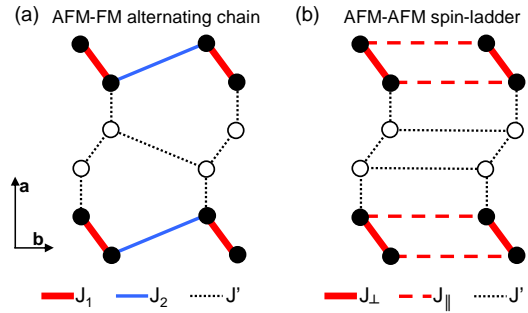}
\caption{\label{fig:mcclstructure} Cu$^{2+}$ $S=1/2$ sites in an
$ab$ plane of DMACuCl$_3$.  Open and closed circles illustrate two
distinct dimer bonds/chains in the LT phase\cite{willet_lowTstructure}.  (a) AFM-FM
HAC scenario with $J_{1}$ and $J_{2}$ shown as solid lines and weaker
exchange, $J^{\prime}$, shown as dotted lines.   (b) AFM-AFM SL scenario
with $J_{\perp}$ and $J_{\parallel}$.
Lattice vectors correspond to HT monoclinic
parameters \cite{willetstructure}.}
\end{figure}

We report new inelastic neutron scattering (INS) results for $\mathrm{(CH_{3})_{2}NH_{2}CuCl_{3}}$ [$\mathrm{DMACuCl}_3$ or MCCL].  We use state of the art linked-cluster calculations \cite{zheng2006,expansions2} and
bulk measurements to show $\mathrm{DMACuCl}_3$ contains a realization of the AFM-FM HAC with $\alpha =  -0.92(4)$.  Moreover, in contrast to results on existing AFM-AFM HACs \cite{gxucoppernitrateprl,blake1997}, there is evidence for spatially extended entanglement of the principle dimers via the FM coupled spins.

$\mathrm{DMACuCl}_3$ is monoclinic at room temperature
with $\beta = 97.5^{\circ}$ with
copper-halide planes separated along $c$ by
methyl groups, such that magnetic coupling is only
expected in the $ab$-plane as shown in 
Fig.~\ref{fig:mcclstructure}\cite{willetstructure}.
Thermodynamic and structural considerations
were used to identify $\mathrm{DMACuCl}_3$ as an AFM-FM
HAC along the $a$-axis with $\alpha \approx -1$
\cite{gerstein1972,hurley1973,obrien1988}, attracting
interest since other experimental realizations of 
AFM-FM HACs have $\alpha \ll -1$ \cite{nishikawa1998,afmfmchainchicalc}.
Further measurements were interpreted
as independent FM and AFM dimers with evidence for
a transition to long range order below $T=0.9$ K \cite{ajirophysicaB,stoneLT24}. Preliminary INS
in the disordered phase eliminated
both independent dimer and $a$-axis chain models \cite{stoneICNS}.
Recently, it was shown that the low-temperature (LT), $T < 285$ K,
structure is triclinic with two independent
chains along the high-temperature (HT) $b$-axis \cite{InagakiJPhysJpn2005,willet_lowTstructure}, \emph{cf.}
Fig.~\ref{fig:mcclstructure}.
Bond lengths in the LT structure favor HAC 
Cu-halide-halide-Cu or Cu-halide-Cu exchange paths, found to be relevant in other low-d
magnets\cite{stonephccprb,zheludevIpac,turnbullsummary}.

Deuterated samples were obtained by slow
evaporation of a D$_2$O solution of CuCl$_2$ and
(CD$_3$)$_2$ND$\cdot$HCl.  A 2.23 g single crystal mounted in the ($hk0$) plane was examined using the
SPINS spectrometer at NIST configured with $80^{\prime}$
collimation before and after the sample and a cooled Be filter in
the scattered beam.  A flat PG(002) analyzer selected
scattered neutrons at 5 meV.
Constant wave-vector, $\mathbf{Q}$, scans were performed
at $T=1.8$ K indexing $\mathbf{Q}$ in the HT
monoclinic notation with LT lattice constants,
$a$=12.05 \AA~and $b$=8.43 \AA \cite{willetstructure}.

\begin{figure}
\centering\includegraphics[scale=0.80]{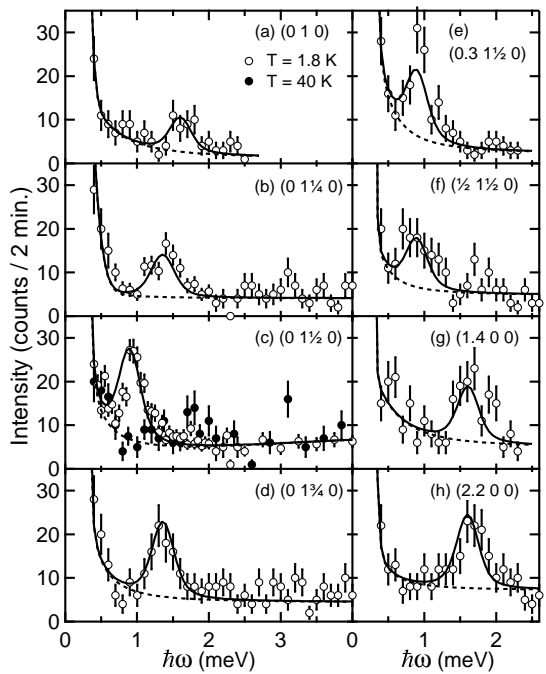}
\caption{\label{fig:mcclscans} $T=1.8$ K constant $\mathbf{Q}$ scans
of DMACuCl$_3$.  (c) includes data at $T=40$ K.
Lines are based upon a global
fit to the AFM-FM HAC model described in the text.  Dashed
lines are fitted backgrounds.}
\end{figure}

Representative scans are shown in Fig.~\ref{fig:mcclscans}.
Energy transfers were scanned up to $\hbar\omega \approx 4.5$ meV.
A single peak disperses along
$k$ between $0.95(2) \leq \hbar\omega \leq 1.66(2)$ meV as shown in Fig.~\ref{fig:mcclscans}(a)-(d).  The mode is only
weakly dispersive along
$h$ at the bottom (Fig.~\ref{fig:mcclscans}(e) and (f)) and top (Fig.~\ref{fig:mcclscans}(g) and (h))
of the band with $\approx 0.2(1)$ meV dispersion along
($h$ 1.5 0).  Intensity reduction
for temperatures large
compared to the energy scales of excitations
indicates a magnetic origin, \emph{cf}. Fig.~\ref{fig:mcclscans}(c).
The intrinsic peak width is small compared to the
instrumental resolution of $\approx 0.25$ meV FWHM.  We do not see any reasonable evidence for gapless excitations or magnetic continuum scattering.

\begin{figure}
\centering\includegraphics[scale=0.80]{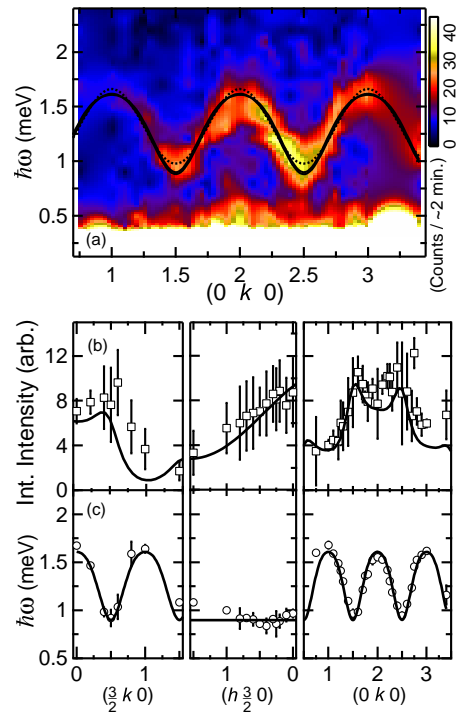}
\caption{\label{fig:mccldisp}(a) Scattering intensity
\emph{vs.} $\hbar\omega$ along $(0k0)$.  Solid lines are fits to dispersion based upon
linked-cluster models.  Dotted
line is dispersion based upon perturbation theory for weakly coupled
dimers.  (b) Integrated intensity \emph{vs.} $\bf{Q}$.
(c) Dispersion based upon Gaussian peak approximation.
Lines in (b) and (c) are based upon a global fit
to the AFM-FM HAC model described in
the text.}
\end{figure}

We fit the constant $\mathbf{Q}$ scans with
Gaussian peaks to determine dispersion and intensity
variation in the $(hk0)$ plane as summarized in
Fig.~\ref{fig:mccldisp}.
Lack of dispersion along $h$
excludes the proposed $a$-axis as the chain axis, and
significant dispersion along $k$ excludes
non-interacting dimer models.
The $k$ dispersion has periodicity $2\pi$ in reduced units, a
quantum effect
showing a lack of AFM order.
The FM-FM HAC is ruled out by the absence of a second dispersive
mode \cite{Huang1991}.
Notably, the band maximum[minimum] is
at $Q =  2 m \pi$[$Q= (2m + 1) \pi$] for integer $m$.
This is unusual and rules out a HAC with $\alpha > 0$.  
Models that can produce such dispersion include the AFM-FM HAC and the AFM-AFM spin-ladder (SL) with the HAC favored
due to structural arguments.

\begin{table}[t]
\begin{tabular}{c|r|r}
  \hline
  model (method) & $J_1$ or $J_{\perp}$ (meV)& $J_2$ or $J_{\parallel}$ (meV) \\
  \hline
  AFM-FM HAC (E)  & 1.406(8) & -1.30(5) \\
  AFM-FM HAC (PT) & 1.319(8) & -0.68(2) \\
  AFM-AFM SL (E)  & 1.194(6) &  0.376(5) \\
  AFM-AFM SL (PT) & 1.319(8) &  0.34(1) \\
  \hline
\end{tabular}
\caption{Resulting exchange parameters from
fitting dispersion to AFM-FM HAC and AFM-AFM SL models using
linked-cluster expansions (E) and perturbation theory (PT).}
\label{tab:exchangetable}
\end{table}

The dispersion for the HAC and SL models can be
calculated using perturbation theory, with an
identical form to first order, $\hbar\omega(\mathbf{Q})=
\epsilon_1 + \epsilon_2 \cos(\mathbf{Q}\cdot\mathbf{u})$
\cite{abharris,barnesafmchain,barnes_prb1993,HidaPRBs1992}, where $\epsilon_1=J_1[J_{\perp}]$ and $\epsilon_2=\frac{|J_2|}{2}[J_{\parallel}]$ for
the HAC [SL] models.  It is
more accurate to use state of the art linked-cluster expansion methods \cite{zheng2006,expansions2}.  Parameters obtained from fits of the data to these models are summarized in Tab.~\ref{tab:exchangetable}.
Dispersion alone can not differentiate between the HAC and SL models. Solid lines in Fig.~\ref{fig:mccldisp}(a) show the fit to the linked
cluster calculation and the dotted line a fit to the perturbation
theory.  Perturbation theory is clearly inadequate to describe
the AFM-FM HAC as evidenced by the large difference in fitting
parameters relative to the linked cluster method.  Although fits to the AFM-FM HAC dispersion characterize the magnitude of FM and AFM exchange, this comparison does not include structural information determining which bond in Fig. 1(a) is FM or AFM.

\begin{figure}
\centering\includegraphics[scale=0.75]{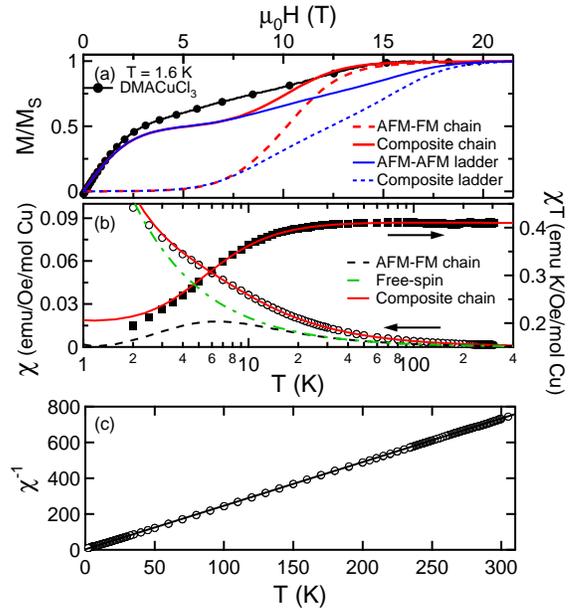}
\caption{\label{fig:mcclmagnetization}(a) Pulsed field
magnetization of DMACuCl$_3$ powder at
$T = 1.6$ K.  Lines are QMC calculations based upon AFM-FM
HAC and AFM-AFM SL models.  Solid lines, labeled as
composite chain or composite ladder, include a free-spin contribution.
Symbols for data are plotted every 1000 points.  
(b) Single crystal magnetic susceptibility, $\chi(T)$,
and $\chi(T)T$ for $\mu_0 H=0.05$ Tesla with $H\parallel a$.
Lines are the composite model AFM-FM HAC polynomial
fit discussed in the text.  Broken lines are the $\chi(T)$ contribution
of the AFM-FM HAC (dashed black) and free-spin (dash-dot green)
components of the composite model.  (c) $\chi^{-1}(T)$ based upon
data in (b).  Solid line is extrapolated linear fit described in
the text.}
\end{figure}

In principle, high field INS can be used to determine
absolute values of exchange parameters assuming they are field
independent \cite{coldea2002prl}.  Unfortunately, to characterize DMACuCl$_3$ accurately in this manner would require INS
at magnetic fields well over $\mu_0 H=20$ T; not possible at
the present time.  However, it is practical to examine
high field bulk thermodynamic properties.
Figure~\ref{fig:mcclmagnetization}(a) shows pulsed field
magnetization of DMACuCl$_3$ up to 20 T at $T=1.6$ K.  The
upper critical field where all spins are FM aligned is
in the vicinity of 16 T.  We also plot
Quantum Monte-Carlo (QMC)
calculations of the magnetization for 100 spins using
exchange constants determined from cluster-expansion fits of $\hbar\omega(\mathbf{Q})$ \cite{alps}.  The AFM-FM HAC model reproduces the upper
critical field indicating that the determined energy scales are more appropriate, but
neither model accounts
for the linear magnetization at low-field.

Recall that the LT structure has
two types of bonding, \textit{i.e.} two sets of exchange parameters for quasi-1d
chains along the monoclinic $b$-axis.
However, our
INS measurements indicate only a single mode.  The energy
scale of the second set of quasi-1d chains is either
much higher, lower or identical to that of the
excitation we observe.  Higher energy excitations would manifest
a higher energy gap in the spectrum, not seen in
thermodynamic measurements.  It is most reasonable
that these excitations are at lower energies.  The present
experiment sets an upper limit of $J^{\prime}\approx 0.2$ meV for this energy scale, indistinguishable from the incoherent elastic scattering.
In the absence of a measured exchange
for the weakly coupled chains,
we model these moments as free-spins.  This is reasonable given
the temperature scale of the magnetization measurements.
A conjecture of a free-spin
contribution was also made in Ref.~\cite{watsonthesis} to
account for magnetization measurements.  We therefore calculate
$M(H)$ as a composite model with half the moments included
as free-spins.  This accounts for the initial
linear magnetization. As seen in figure 4(a), the composite
AFM-FM HAC model calculation is a much better description of the the
measurement than the composite AFM-AFM SL model.  When $M(H)$ is calculated using parameters derived from perturbation theory the situation is similar, but as expected the overall agreement with the data is not as good for either model.

The calculations illustrated in Fig.~\ref{fig:mcclmagnetization}(a) are
performed with no adjustable parameters.  The AFM-FM HAC model accounts very well for both the low field and high field limits with deviations in the vicinity of 5 to 10 T.  This may be a consequence of a field-induced collective ordered phase \cite{stoneLT24} that is not accounted for in the calculation.   Willet \textit{et al.} used finite chain calculations to obtain parameters that reproduce $M(H)$ within a two chain model\cite{willet_lowTstructure}.  However, the exchange parameters they propose are incompatible with the INS data reported here, and can be ruled out by the observation of only a single excitation. 

Figure~\ref{fig:mcclmagnetization}(b) shows temperature dependent magnetic susceptibility, $\chi(T)$.  The solid line is a fit using the AFM-FM HAC:free spin composite model, with $\chi(T$) for the HAC calculated using a polynomial representation derived from finite-sized scaling \cite{afmfmchainchicalc}.  The parameters $J_1 =  0.973(4)$, $J_2 = -1.23(5)$ meV and $g=2.096(1)$ are reasonably consistent with INS results although clearly the latter provides a more direct measure of the exchange parameters.  Figure~\ref{fig:mcclmagnetization}(c) shows $\chi^{-1}(T)$.  A linear fit for $120<T<270$ K leads to a value for the Curie-Weiss temperature, $\Theta =-0.9(1)K$. Linear fits to the composite model calculations of $\chi^{-1}(T)$ yield $\Theta =-0.4(3)K$ for the AFM-FM HAC, and $\Theta =-2.4(7)K$ for the AFM-AFM SL confirming the chain model is in better accord with the data.  

Neutron scattering intensities for a straight HAC can be calculated using the coupled-cluster expansions~\cite{zheng2006}.  Comparison to these
using our HAC exchange parameters and noting the broad wave-vector dependence of the integrated intensity, \emph{cf.} Fig.~\ref{fig:mccldisp}(a) and (b), the short bond in Fig.~\ref{fig:mcclstructure} is likely AFM.
However, the bonding geometry in DMACuCl$_3$ is complicated by in-plane zig-zag and small out of plane components.   We incorporate bond vectors
within a single mode approximation (SMA) \cite{gxucoppernitrateprl,stonephccprb} to calculate the dynamic spin correlation function.  The SMA is justified by the
observation that the INS data show a single peak with
no appreciable additional intensity, and
that the calculated spectral weight is dominated by single particle
excitations for $\alpha \approx -1$ \cite{zheng2006}.
The SMA to the scattering intensity is
\begin{eqnarray}
\label{SqSMA}
 \tilde{{\cal I}}_m ({\bf Q}, \hbar\omega) & \propto &  
 \frac{|F(Q)|^2}{\hbar\omega({\bf Q})} \sum_{{\bf d}} J_{{\bf d}}\langle{\bf S}_0 \cdot {\bf S_{\bf d}}\rangle [1-\cos{({\bf Q} \cdot {\bf d})}] \nonumber \\
& & \delta(\hbar\omega - \hbar\omega({\bf Q})).
\end{eqnarray}
$\langle{\bf S}_0 \cdot {\bf S_{\bf d}}\rangle$ is the spin correlation function, $\bf d$ is a bond vector, and the sum is over the AFM and
FM bond, ${\bf d}_{1}$ and ${\bf d}_{2}$.
We use 74 constant ${\bf Q}$ scans ($\hbar\omega \geq 0.5$ meV) in a global fit to Eq.~\ref{SqSMA}
convolved with the instrumental resolution function and a
dispersion based upon the AFM-FM HAC cluster-expansion using parameters
in Tab.~\ref{tab:exchangetable}.  Parameters include an overall scaling factor and the ratio
$A=\langle{\bf S}_0 \cdot {\bf S}_{{\bf d}_{2}}\rangle / \langle{\bf S}_0 \cdot {\bf S}_{{\bf d}_{1}}\rangle$.  The results reproduce well the dispersion and intensity modulation and are shown as solid lines in Figs.~\ref{fig:mcclscans} and~\ref{fig:mccldisp}(b,c) with $A=0.26(4)$. Allowing exchange constants to be fit parameters results in $J_1 = 1.399(6)$ and $J_2 = -1.07(3)$ meV, and $A=0.27(4)$.

The SMA has been applied to the AFM-AFM HAC, with the result that $\langle{\bf S}_0 \cdot {\bf S}_{{\bf d}_{1}}\rangle$ is consistent with the isolated dimer expectation of $-3/4$, and the empirical value for $\langle{\bf S}_0 \cdot {\bf S}_{{\bf d}_{2}}\rangle=0$\cite{gxucoppernitrateprl}.  The significance of this is discussed by Brukner \emph{et al.} \cite{brukner}, who shows that the intra-dimer correlation is a manifestation of quantum entanglement in the bulk system.  Entanglement is localized to within one dimer for the AFM-AFM HAC.  In contrast for DMACuCl$_3$, the inter-dimer spin correlation is clearly non-zero.  This is expected since as alpha tends to $-\infty$ the FM coupled spins can evolve towards a composite $S=1$ entity, with a $T=0$ spin-spin correlation in the Haldane chain extending over several lattice spacings \cite{regnault94}.  Since the quantum correlation becomes spatially extended, adjacent dimers in the AFM-FM HAC can be thought of as more entangled than those in the AFM-AFM HAC.  Notably, fitting the SMA for a purely AFM SL to the present data yields $A\approx 0$, indicating suppression of interdimer entanglement for purely AFM coupled dimers in that case as well.

In summary, high resolution INS, combined with high field magnetization and susceptibility measurements show that DMACuCl$_{3}$ is a quasi-1d AFM-FM HAC with $\alpha \approx -1$, intermediate between weakly coupled dimer and Haldane regimes.  The exchange parameters are determined from the dispersion using the results of coupled cluster series expansions.  Applying the SMA to analyze the scattering intensity shows that the inter-dimer spin correlation is significant, indicating that the AFM-FM HAC may be a model system for studying the effect of spatially extended quantum entanglement on bulk properties.

We acknowledge discussions with M. Meisel, A. Zheludev, T. Barnes,
R. R. P. Singh and W. Zheng.   DM acknowledges B. Lake for
drawing his attention to DMACuCl$_3$.  
Research sponsored by the Division of Materials Sciences and
Engineering, Office of Basic Energy Sciences, U.S. Department
of Energy, under contract DE-AC05-00OR22725 with Oak Ridge
National Laboratory, managed and operated by UT-Battelle, LLC.
This work utilized facilities supported in part by the National
Science Foundation under Agreement No. DMR-0454672.  The NHMFL is
supported by the DOE, NSF and Florida State University.

\end{document}